\input epsf     
\documentstyle[mu95,12pt]{article}
\def\beq{\begin{equation}}
\def\eeq{\end{equation}}

\def\beqn{\begin{eqnarray}}
\def\eeqn{\end{eqnarray}}
\def\={\ =\ }
\title{PRECISION MEASUREMENTS AT A MUON COLLIDER}
\author {
S. Dawson  \\
Physics Department \\
Brookhaven National Laboratory\\
  Upton, New York  11973                 \vspace{.5pc} \\
}
\begin{document}
\maketitle
\vspace{4pc}
\medskip
\aabstract
{We discuss the potential for making precision measurements
of $M_W$ and  $M_T$ at a muon collider and the
motivations for each measurement.
A comparison is made with the precision measurements
expected at other facilities.   The measurement of the
top quark  decay width
is also discussed.
     }

\section{INTRODUCTION}
 A $\mu^+\mu^-$ collider with high luminosity and narrow beam
spread offers the possibility of performing high precision measurements of
fundamental masses and decay widths occurring in the Standard
Model and in some extensions of the Standard Model.  We discuss precision
measurements of the $W$ mass and  the top quark mass and width.
Measurements of the Higgs boson mass and width, both in the Standard
Model and in SUSY models are discussed in Ref. [1].
  We pay
particular attention to the motivation for making each precision
measurement and the experimental precision which is necessary in order
to test the theoretical consistency of the Standard Model
or to verify the existence of new physics.

Each of these precision measurements depends on knowing the relevant
energy and building a storage ring to maximize the luminosity at that
energy.
The mass measurements of the $W$ and  top quark
  are  made
by scanning the threshold energy dependences  of the cross sections.  The
threshold energy dependences will be smeared by radiation from the
initial state particles, limiting the precision of the
measurements.  Because the muon is much heavier than the electron,
there will be less initial state
radiation and the beam energy resolution
may be better in a muon collider than in an electron collider,
leading to the possibility of more precise measurements.

\section{ MEASUREMENT OF THE  $W$ MASS}

A precision measurement of the $W$ mass is of fundamental
importance to our understanding of the Standard Model.  Combined with
a precision measurement of the top quark mass,
 the consistency of the Standard
Model can be checked since the $W$ mass is predicted as a function
of the top quark mass.

The current world average on the $W$ mass is obtained by combining data
from UA2, CDF, and D0: [2]
\beq
M_W=80.23\pm .18~GeV.
\eeq
With more data from CDF and D0, both
the systematic and statistical errors will decrease and it has
been estimated [2] that with $100~pb^{-1}$ it will be possible to
obtain:
\beq
\Delta M_W^{\rm~Tevatron} \sim 110 \pm 20~MeV,
\eeq
while  $1000~pb^{-1}$ will give
\beq
\Delta M_W^{\rm ~Tevatron}\sim 50\pm 20~MeV,
\eeq
where the first error is statistical and the second is systematic.

At LEP-II, the error on $M_W$ can  be reduced still further.
  There are two general strategies for obtaining a mass measurement.
The first is to reconstruct the decay products of the $W$, while the
second method is to measure the excitation curve of the $W$ pair
production cross section as the energy is varied. Both methods give
approximately the same precision.   By reconstructing the
$W$ decay products  with
$500~ \hbox{pb}^{-1}$ ($3$ years running) at $\sqrt{s}=190~MeV$[2],
\beq
\Delta M_W^{\rm~LEP-II}  \sim 40~MeV. \eeq
The precision is ultimately limited by the knowledge of  the
beam energy, $\Delta E^{\rm~beam}\sim 20~MeV$.

It is possible that a muon collider could obtain a more precise
measurement of $M_W$ than is possible at LEP-II.  We will discuss
the design restrictions   on a muon collider
in order  to make this the case. The
beam spread at a lepton  collider can be roughly assumed to have
a Gaussian energy resolution with a rms deviation,[1]
\beq
\delta\sim 60~MeV\biggl({R\over .06\%}\biggr)\biggl({\sqrt{s}\over 2 M_W}
\biggr) ,
\eeq
leading to an energy resolution smaller that the $W$ decay width.
A typical parameter for a high energy $e^+e^-$ collider is $R=1$,
 while a lower value is envisioned for
a muon collider, (say $R\sim .06~\%$),
due to the fact that a muon collider will have less
initial state radiation (ISR) than an $e^+e^-$ collider.  Here
we investigate the requirements on $R$ in order for a  measurement
of $M_W$ to be made at a muon collider  which will improve on the precision
expected at  LEP-II.

The procedure is to study the shape of the $l^+l^-\rightarrow W^+W^-$
cross section as a function of the center-of-mass energy, $\sqrt{s}$,
and to fit a theoretical expectation to the cross section.  The many
theoretical effects which must be included are discussed by Stirling
[3]  and
we follow his discussion closely.

We compute the cross section for off-shell $W$ pair production
including Coulomb effects  as,[3]
\beq
\sigma_a(s)(l^+l^-\rightarrow W^+W^-)
=\biggl[ 1+\delta _C(s)\biggr]  \int ds_1 \int
ds_2 \rho(s_1)\rho(s_2)\sigma_0(s,s1,s2),
\eeq
where
\beq
\rho(s)={\Gamma_W\over \pi M_W}\biggl[{s\over (s-M_W^2)^2
+s^2\Gamma_W^2/M_W^2}\biggr]
\eeq
 and $\sigma_0(s,s_1,s_2)$ is the Born cross section for
producing a $W^+W^-$ pair with  $W^\pm$ energies  $\sqrt{s_1}$
and  $\sqrt{s_2}$.
(Electroweak radiative corrections are negligible in the
threshold region.) This procedure defines what we mean by
the $W$ mass and is in fact the same definition as used in
LEP measurements.  The Coulomb corrections are included in
the factor $\delta_C(s)$ and arise from the fact the
$W^+W^-$ cross section diverges as $1/v$ at threshold.
The analytic expression for $\delta_C(s)$ can be found in
Ref. [4].

Finally, the corrections due to initial state radiation, which
are sensitive to the lepton mass, $m_l$,
 must be included.  The ISR is the
first correction which differentiates between an electron and a muon
collider.  These corrections
 are included with a radiator function $F(x,s)$ (given
in Ref. [5]) to obtain our final result for the $W^+W^-$ pair production
cross section:
\beq
\sigma(s)(l^+l^-\rightarrow
W^+W^-)={1\over s}\int d s^\prime F(x,s) \sigma_a(s^\prime)
\eeq
where $x\equiv 1-{s^{\prime}\over s}$ and
\beq
F(x,s)\sim t x^{t-1} + \biggl({x\over 2}-1\biggr) t + ...
\eeq
with
$t={2\alpha\over\pi}\biggl[ \log({s\over m_l^2})-1\biggr]$ .
We see that the  ISR is potentially much larger at an $e^+e^-$ machine
than at a  muon collider.
The various contributions to $l^+l^-\rightarrow W^+W^-$ are shown in Fig. 1.
Near threshold,  $\sqrt{s}\sim 2 M_W$,
 the cross section rises rapidly with
 $\sqrt{s}$.  As expected,
there is less reduction of the cross section due to
ISR at a muon collider than at an electron collider.
In Fig. 2, we see that the cross section is  very sensitive to
the precise value of $M_W$ assumed.  It is straightforward to find
the statistical error for a given efficiency, $\epsilon$, and
luminosity, $\int {\cal L}$,[3]

\beq\Delta M_W^{stat}=
{1\over \mid{d\sigma\over dM}\mid}\sqrt{{\sigma\over\epsilon
\int {\cal L}}}.
\eeq
{}From  Fig. 3, we find  that the minimum statistical error
occurs at $\sqrt{s}\sim 2 M_W$ (where the cross section has
the steepest dependance on energy).
The statistical error is obviously not  much different at
a muon collider than at an electron collider, so our
results correspond to those of Ref. [3].
That this must be the case can be seen from Fig. 1; at
threshold, the effects of initial state radiation are small.
At the minimum:
\beq
\Delta M_W^{stat}\sim 90~MeV\biggl[{\epsilon \int {\cal L}\over
100~pb^{-1}}\biggr]^{-1/2}.  \eeq
If it were possible to have $1~ fb^{-1}$ concentrated at
$\sqrt{s}\sim 2 M_W$ with an efficiency $\epsilon=.5$, then
a muon collider could find $\Delta M_W^{stat}\sim 40~MeV$!
Unfortunately, the luminosity of a muon collider decreases rapidly
away from the design energy, so $1~fb^{-1}$
at $\sqrt{s}\sim 160~GeV$  is  probably an
unrealistic goal.

It is also necessary to consider the systematic error, which
is primarily due to the uncertainty in the beam energy.  From Fig. 2,
changing the beam energy is equivalent  to a shift in $M_W$,
$\Delta M_W^{sys}
 \sim \Delta E^{beam}$.
Therefore,  to  obtain a measurement with
the same precision as  LEP-II, a muon collider must be designed
with
 $\Delta E^{beam}\sim 20~MeV$, which corresponds to $R
\sim .02\%$.

The bottom line is that
a $\mu^+\mu^-$  collider must have on the order of $1~fb^{-1}$ at
$\sqrt{s}\sim 2 M_W$ and an extremely narrow beam spread,
$R\sim .02~\%$, in order to be competitive with LEP-II for
a measurement of  $\Delta M_W$.

\section{ PRECISION MEASUREMENT OF THE TOP QUARK MASS AND WIDTH}

A muon collider will also be able to obtain a very precise
measurement of the top quark mass, as  has been discussed in detail
by Berger.[6]   Here, we concentrate on the physics motivations for
making a precision measurement of the top quark mass.

Since at the moment we have no firm prediction for the top
quark mass, a precision measurement of $M_T$ is not particularly
interesting in itself.  However, when combined with a precision
measurement of $M_W$, it tests the
consistency of the  Standard Model.
This is because the prediction for
 $M_W$  in the Standard Model  depends on $M_T$,[7]
\beq M_W^2=M_Z^2\biggl[ 1 -{\pi\alpha\over \sqrt{2} G_\mu M_w^2
(1-\Delta r)}
\biggr]^{{1\over 2}}
\eeq
with
$\Delta r\sim {M_T^2\over M_W^2}$, ($\Delta r$ also depends
logarithmically on the Higgs boson  mass).
In Figure 4, we show the relationship between $M_T$ and $M_W$ in
the Standard Model (where we have assumed $M_H=100~GeV$
and included only contributions  to $\Delta r$
which depend quadratically on
the top quark mass).
For $M_T=175~GeV$,
a measurement of $M_W$ to
$\Delta M_W=40~MeV$ requires a measurement
of $M_T$ to $ \Delta M_T=6~GeV$
 in order to check the consistency of the Standard Model,
while
$\Delta M_W=20~MeV$ requires $ \Delta M_T=3~GeV$.
Given the expected precision on $\Delta M_W$ at LEP-II,
it is clear that there is
 no motivation for a more precise measurement of $M_T$ than
several GeV.

A precise measurement of $M_T$ and $M_W$  also gives some
information
on  the Higgs mass.
 For example, if $\Delta M_W=40~MeV$, $\Delta M_T=4~GeV$
 and the true value of $M_H$ were $100~GeV$, then one could
deduce from the electroweak measurements  that at the $1~\sigma
$ level,
 $50 < M_H < 200~GeV$.[2]

{}From the Tevatron, we will have $\Delta M_T\sim 8~GeV$ with
  $100~ pb^{-1}$ and
$\Delta M_T\sim \pm 4~GeV$ with $1000~pb^{-1}$.[2]  The LHC experiments
are designed such that with 1 year of  running,
$\int{\cal L}= 10~fb^{-1}$, a value on  the order  of,
$ \Delta M_T\sim 3~GeV$ will be obtained.[2]
 In contrast, the values which would be obtained from a muon collider [6]
\beq
\Delta M_T^{\mu^+\mu^-}\sim 300~MeV
\eeq
and from an electron collider [8]
\beq
\Delta M_T^{e^+e^-}\sim 520~MeV\eeq
are considerably more precise.

A precision measurement of the top quark width conveys
significantly more information than a precision measurement
of the mass.  This is because the width is sensitive through
loop effects  to new
particles contained in extensions of the Standard Model.
Single top production at the Tevatron will measure the
top quark width to roughly,[9]
\beq
{\Delta \Gamma_T^{\rm{Tevatron} }
\over
\Gamma_T}\sim .3,
\eeq   while a $500~GeV$ $e^+e^-$ collider might
obtain[8]
\beq
{ \Delta \Gamma_T^{e^+e^-}\over \Gamma_T}\sim .2 \quad .
\eeq
  Presumably,  a $\mu^+\mu^-$  collider will do even better.
Such measurements will be capable of limiting the low mass particle
spectrum of  supersymmetric models and it would be interesting to have
a systematic comparison of the capabilities of an $e^+e^-$ and
$\mu^+\mu^-$ collider and the corresponding limits on SUSY particles.

\section{REFERENCES}

\noindent 1.  See, for example, V.~Barger, M.~Berger, J.~Gunion, and
 	T.~Han,  Phys. Rev. Lett. {\bf 75} (1995) 1462,
      hep-ph/9504330  ;  V.~Barger, contribution to this workshop.

\noindent 2.   F.~Merritt, H.~Montgomery, A.~Sirlin, and M.~Swartz,
{\it Precision Tests of Electroweak Physics},
	report of the DPF Committee on Long Term Planning, 1994.

\noindent 3. W.~Stirling, DTP-95-24 (1995), hep-ph/9503320.

\noindent 4.  V.~Fadin, V.~Khoze, A.~Martin, and W. Stirling,
Phys. Lett. {\bf B363} (1995) 112, hep-ph/9507422.

\noindent 5. F.~Berends, in {\it Z~Physics~ at~ LEP~I}, CERN Yellow
Report No. 89-08, Geneva, 1989, Vol. 1, edited
 by G.~Altarelli, R.~Kleiss, and  C.~Verzegnassi.

\noindent 6.  M.~Berger, presented at {\it International~Symposium~
on~Particle~Theory~and Phenomenology}, Ames, Iowa (1995),
hep-ph/9508209.

\noindent 7.  W.~Marciano, Ann. Rev. Nucl. Part. Sci {\bf 41} (1991) 469.

\noindent 8. P.  Igo-Komines, in {\it Proceedings~of ~the~Workshop
{}~on ~Physics~and~Linear~$e^+e^-$~Colliders}, Waikoloa,
Hawaii, 1993, edited by F.~Harris, S.~Olsen, S.~Pakvasa,
and X.~Tata (World Scientific, Singapore, 1993);
 K.~Fujii, T.~Matsui, and Y.~Sumino,
Phys. Rev. {\bf D50} (1994) 4341.

\noindent 9.  D.~Carlson and  C.P.~Yuan, presented
at {\it International~Symposium~on ~Particle~Theory~ and~
Phenomenology}, Ames, Iowa (1995),  hep-ph/9509208.

\begin{figure}[t]
\epsfxsize=4.5in
\epsfysize=4.5in
\centerline{
\epsffile {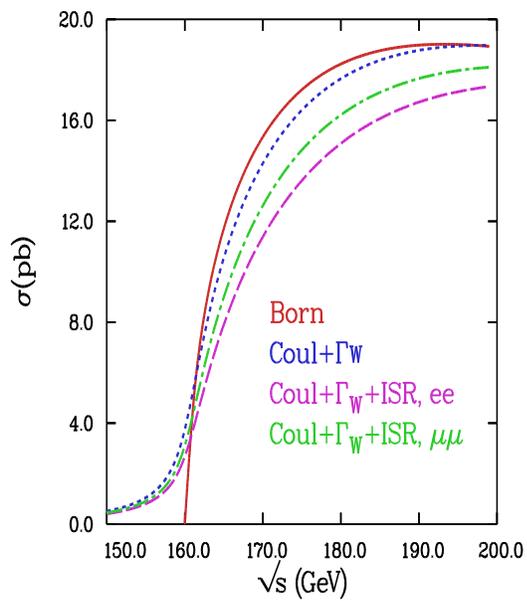}\hfil
}
\vskip 1.0in
\caption[]{Contributions to the $l^+l^-\rightarrow W^+W^-$ cross
section.  The solid curve includes only the tree level cross section,
while the dotted curve includes the Coulomb and finite $W$ width
effects.  The long-dashed and dot-dashed curves include the effects
of ISR for $e^+e^-$ and $\mu^+\mu^-$ colliders, respectively.  }
\end{figure}
\vskip 2.in

\begin{figure}[t]
\vskip .6truein
\epsfxsize=4.5 in
\epsfysize=4.5in
\centerline{\epsffile {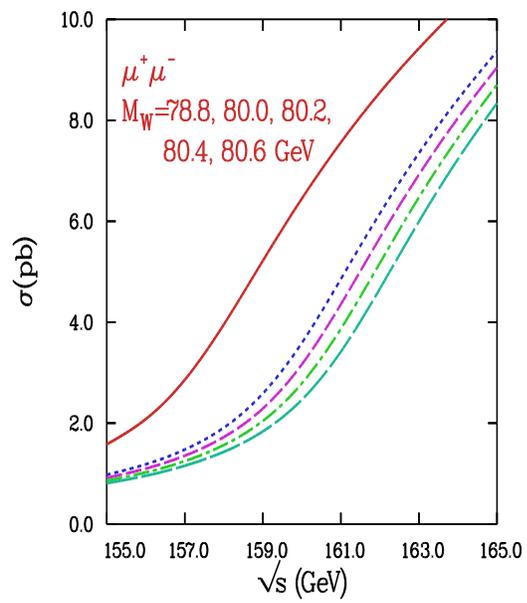}}
\vskip 1.in
\caption{Cross section for $\mu^+\mu^-\rightarrow W^+  W^-$.
The solid curve has $M_W=78.8~GeV$. }
\end{figure}
\vskip 1in

\begin{figure}[t]
\vskip .6truein
\epsfxsize=4.5in
\epsfysize=4.5in
\centerline{\epsffile {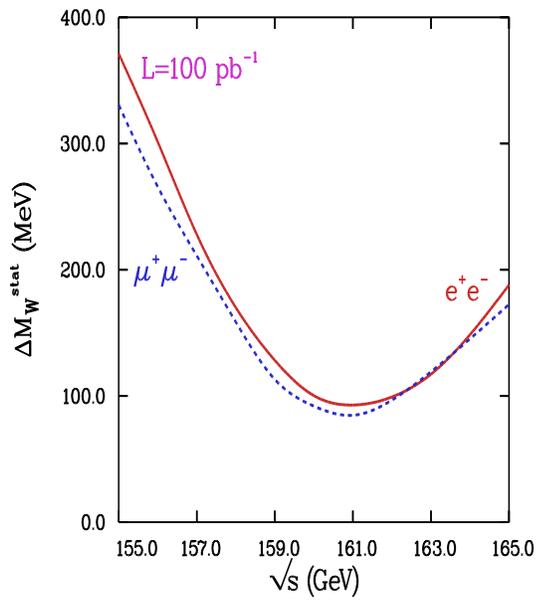} }
\vskip 1in
\caption{Statistical error on $M_W$ from
$l^+ l^-\rightarrow W^+W^-$  from an absolute
measurement of the rate with an integrated luminosity,
${\cal L}=100~pb^{-1}$. }
\end{figure}
\vskip 1in

\begin{figure}[t]
\epsfxsize=2.5in
\epsfysize=2.5in
\centerline{\epsffile{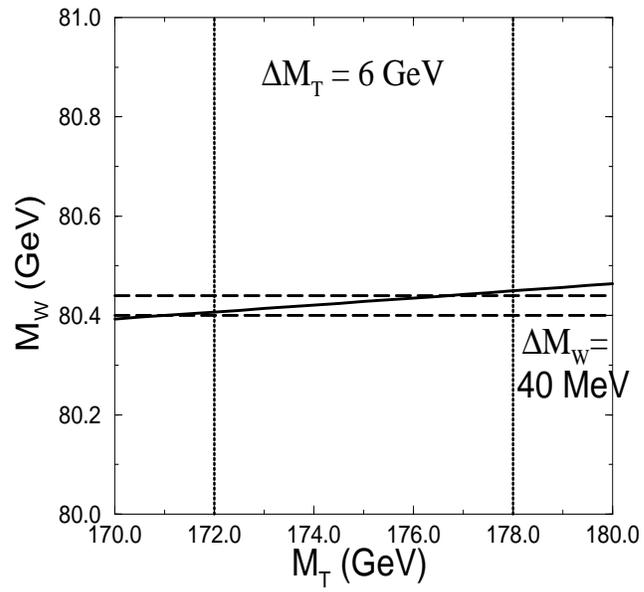}}
\vspace{1.5in}
 \caption{Dependence of the predicted $W$ mass on the top quark
mass in the Standard Model. }
\end{figure}

\vskip .5in

\end{document}